# COMMUNITIES IN AFFILIATION NETWORKS
# WITH ATTITUDINAL ACTORS

Moses A. Boudourides

Department of Mathematics
University of Patras, Greece
mboudour@upatras.gr

## Network Communities within Analytical Sociology

One of the core issues (if not *the* core issue) in analytical sociology is the investigation of the micro-macro link in social phenomena (Hedström & Swedberg, 1998, Hedström, 2005). Analytical sociology endorses a social mechanism-based type of explanation according to which causal relationships between social phenomena at the macro-level are explained through the patterns of individuals' action and interaction at the micro-level. This explanatory scheme is clearly illustrated by the concatenation of a number of causal mechanisms encoded in the so-called *Coleman's boat* (Coleman, 1990). In Coleman's architecture, the first causal mechanisms bridging a certain macro-level phenomenon X (the explanans) with the relevant underlying micro-level processes are *situational mechanisms*, through which social structures constrain individuals' action and cultural settings shape individuals' motives and dispositions (desires, beliefs etc.). Subsequently, individuals develop their action and interaction at the micro-level through some *action-formation mechanisms*, typically understood as operating on an informal social network, in which network actors are the implicated individuals and network ties correspond to patterns of deployed interactions. Finally, the individuals' social network is lifted onto the macro-level through some *transformational mechanisms*, which generate various intended and unintended social outcomes identifying a certain social phenomenon Y (the explanandum). In summary, according to the typology of Coleman's boat, the causal *macro-level association* of the explanans social phenomenon X with the explanandum social phenomenon Y is established through the concatenation of three general types of social mechanisms: situational, action-formation and transformational mechanisms (cf. Hedström & Ylikoski, 2010).

In our presentation here, we aim to investigate the macro- (or meso-) level association of the following two social entities: *diversity of attributes* and *segregation in communities*. By the former, we understand a differentiated cultural field within social life, which is framed by such qualities or predicates of behavior as values, preferences, beliefs, positions and opinions that we name altogether "*attributes*." By segregation in communities, we refer to the social order that originally separate social groups may coalesce into higher-order or broader associations, which are still differentiated to each other. Such aggregate concrescences of social groups are what we call *communities* (a term to be further elucidated in the sequel).

Undoubtedly, the hypothesis that formation of communities could be causally driven by existing attributes makes perfect sense. For instance, one could think of the homogenizing or cementing role that culture plays in human sociality and, therefore, one could expect that common attributes might smooth out existing frictions, intensities, differences or disparities among social groups by the concrescence of certain overarching and more comprehensive clusters of groups, meant as communities here. However, our intention is to put forward a mechanisms-based analytic explanatory perspective in order to be able to grasp the macro- (or meso-) level association between attributes and communities. For this purpose, first, we are going to postulate that *distributing attributes* over a population of individuals is the situational mechanism that would connect macro-level cultural attributes to the micro-level population of interacting individuals forming a social network. Now, in a social network, there are many (middle-range) mechanisms that may posit or prove the existence of groups of actors. Such mechanisms can show up through various formal methodologies that look into the structural and causal settings of a given (empirical) social network. Apparently, the corresponding social network analyses vary in complexity according to the type of the examined social network. Here, as a first step, we are going to limit ourselves in the consideration of the simple case of *two-mode* or *affiliation networks*. As a matter of fact, we intend to consider an affiliation network composed of *persons* and *groups* (of persons), which are given empirically (e.g., by means of a survey). This is exactly the setting of the renowned "duality" between persons and groups, which was originally formulated by Georg Simmel (1955) and subsequently it was explored analytically in the seminal work of Ronald Breiger (1974). Therefore, what we consider here to be an action-formation mechanism is the mechanism of *network affiliation duality*, with the understanding that the affiliation of persons into groups was produced by the previous mechanism that had carried out the distribution of attributes on persons (with technical details about the formal operationalization of this mechanism to be described in the next section). Finally, the transformational mechanism that assembles groups into communities (of groups) is considered to be the mechanism of the *emergence of clusters* (of both modes) in an affiliation network with attitudinal actors (to be formally clarified in the following two sections). Thus, the overall architecture of social mechanisms that we are going to attend to here in order to explain the macro- or meso-level association of attributes with communities is given in the following figure.

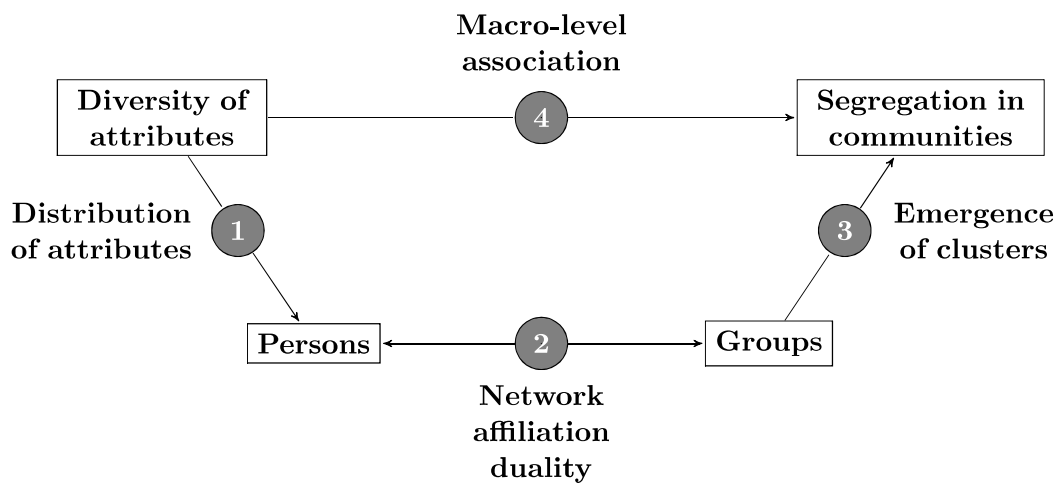

**Figure 1**. The typology of the followed mechanisms.

Having described the social mechanisms called up in our analysis of the causal relationship between attributes and communities, we need to further expound these two notions. First, in what concerns the former, as it will become clear in next section from the point of view of the formal structural methodology that we going to follow, the interference of attributes occurs in two different ways in this mechanism-based causal typology. The first instance is when attributes are conceived as patterns of affiliations that actors happen to have within certain prescribed groupings (of actors).

The second instance is when attributes are construed as patterns of attitudes that actors may exhibit in their transactions. In both instances, attributes operate formally as monadic (not dyadic) actors' characteristics and, thence, the choice of two-mode networks in our analysis (not to mention the scarcity of empirical data sets of one-mode networks, in which actors display a variety of attitudes). However, this is not just a trivial formal simplification that we are obliged to adopt in order to reduce the complexity of the studied network. Substantively, one may evoke what the anthropologist Siegfried Frederick Nadel (1957) used to argue in the past, i.e., "membership roles" correspond completely to "relational roles," a consequence of which is Ronald Breiger's commitment to talk about "two types of social ties: membership and social-relations" (Breiger, 1974, p. 183).[1]

Coming to the notion of communities, we need to stress from the beginning that we are concerned with the relational meaning of the term. Leaving aside technical details (to be discussed in the last two section), what in principle one understands as "community detection" in a social network (with either one- or two-modes) is a partition of the set of actors in certain tight-knit subsets of actors, called communities (Wasserman & Faust, 1994, p. 249). In other words, actors inside a community should be relatively more densely connected to each other than how (relatively sparsely) actors among different communities were connected.

Undoubtedly, to say to which community a social network actor belongs is a type of a group or collective categorization. In any case, a community categorization has nothing to do with what an individual actor may express or imagine of oneself or even on what one's group peers think about oneself. To categorize the community to which an actor belongs is not an issue of an enunciated self-identification, even when an individual actor speculates on existing affinities with other individual actors. For a community (in the relational sense of the term) does not consist of a mere aggregation of persons, conceived as self-reliant subjects, who cogitate about their destiny or recognize themselves in arbitrary or sempiternal terms outside the reality of their social worlds, their everyday practices and life. On the contrary, to categorize an actor's community is an issue of what the actor does in practical contexts, the distinctive ways one acts in mundane situations, how one interacts with or with regards to other actors. Even a person's autonomy (Castoriadis, 1998) or a personal identity, i.e., whatever distinguishes a person from others, are singular characteristics, which develop through a bundle of inter-personal shared significations, embedded in the relational footing of the multiplex ties, structural and cultural, that persons sustain with others in the settings of their common social spaces. This is why Harrison White (2008, p. 157) argues, "communities can be seen as built out of identities rather than persons." Moreover, that "the relationship between community as a complex of social relationships and community as a complex of ideas and sentiments" needs further exploration, as Craig Calhoun (2009, p. 110) was calling forth rather recently, is still another reason why communities and attributes should matter for analytical sociology.[2]

---

[1] Nonetheless, the bias to overrate "affiliational" attributes over attitudinal attributes is still present in certain areas of sociological erudition, as, for example in Doug McAdam's (1986, p. 65) argument of "attitudinal affinity."

## Affiliation Networks

Typically, in social network analysis, a *two-mode network* is a social network comprising two sets of actors (called *modes*) and being endowed with ties only between the two modes (no ties inside a mode). Furthermore, in social network analysis, a particular type of two-mode networks arises when there is a certain inhomogeneity between the two modes. This is the case when the first mode of actors consists of proper individual actors (i.e., actors being individuals or persons) and the second mode represents either collective actors (like groups or organizations) or situations (like events or meetings) to which actors of the first mode belong (either as members of groups or organizations or as participants to events or meetings). Such a two-mode network is usually called an *affiliation network* (Wasserman & Faust, 1994, pp. 29-30).

Now, if, in a social network, there was no predefined (prescribed) differentiation among actors, in the sense that all actors were in a single mode and any actor could possibly form a tie with any other one, then such a social network could be also called a *one-mode network*. According to well-known remark made by Stephen Borgatti and Martin Everett (1997), the difference between one- and two-mode networks can be seen to derive conceptually from the fact that ties in a one-mode network are construed as relations on pairs of actors (i.e., *dyadic attributes* on actors), while in a two-mode network they simply represent (multiple) associations among elements of two distinct modes of actors (i.e., *monadic attributes* on actors, as far as the attribute of an actor in a mode is identified to the actor(s) of the other mode to which the former is associated).

---

[2] The recent article of Shwed & Bearman (2010) represents a good example of how analytical sociology might cope with the relational notion of communities.

As a matter of fact, the previous interpretation is consistent with the distinction between a graph and a hypergraph that is used traditionally in the mathematical field of Graph Theory. Formally (cf., Wasserman & Faust, 1994, p. 147), a *hypergraph* consists of a set of objects and a collection of subsets of objects, in which case each object belongs to at least one subset and no subset is empty (Berge, 1989, p. 1). The objects are called points (or vertices) and the collections of objects are called edges. A *graph* is a special case of a hypergraph in which the number of points in each edge is exactly equal to two. In other words, in a graph, edges are defined as pairs of points (the way, in a one-mode network, ties are dyadic attributes on actors), while, in a hypergraph, edges are defined as arbitrary nonempty collections of points (the way, in a two-mode network, ties are monadic attributes associating actors among the two modes).

Now, if restricted in the case of affiliation networks, according to the previous discussion, an affiliation network is simply a collection of (monadic) attributes on actors (i.e., formally, from the graph-theoretic point of view, it constitutes a hypergraph) - with the understanding that, in an affiliation network, actors' attributes represent such heterogeneous associations, as when actors are members of groups or organizations or they are participating in events or meetings etc.

However, what we intend to examine here is the case of affiliation networks, in which, besides actors' attributes defining an affiliation network (let us call them actors' "*affiliational*" attributes or simply actors' *affiliations* and always keep implicit the fact that they are monadic attributes, not dyadic ones), we have an additional category of actors attributes that (as it was stated in previous section and it will become further clear from the context of the empirical data that we are going to analyze in the last section) we will call actors' "*attitudinal*" attributes or simply actors' *attitudes* (again implicitly considered monadic, not dyadic).

Obviously, from a formal point of view (beyond any particular signification of network terms), an affiliation network with actors' affiliations *and* actors' attitudes is again an affiliation network with regards to a single (but heterogeneous) monadic attribute on actors, which associates actors to both their affiliations and their attitudes. As a hypergraph, we have now two types of edges defined on the same set of points (actors), one corresponding to edges formed by the pattern of actors' affiliations and another one corresponding to edges formed by the pattern of actors' attitudes; clearly, the merger (formally, the union) of these two collections of points is still a collection of points which corresponds to the hypergraph of the original affiliation network augmented by actors' attitudes.[3]

**Community Detection**

The community detection methodology that we are going to apply here is the so-called "modularity maximization technique." We are concerned with the case that the social network is represented by a (undirected simple) graph $G$ on a set of vertices (or points) $V$. Let $n$ be the total number of vertices such that a pair of vertices may form multiple edges (i.e., the graph is weighted) and let $m$ be the total number of edges in the graph. Furthermore, let $A = \{A_{ij}\}$ be the adjacency matrix of the graph (which means that $A$ is a symmetric matrix of order $n \times n$ with entries being non-negative integers such that $\sum_{i, j = 1, ..., n} A_{ij} = 2m$). Now, let $\mathbf{C} = \mathbf{C}(G) = \{C_1, C_2, ..., C_c\}$ be a partition of the set of vertices $V$ of the graph (dividing) $G$ into $c$ subsets $C_k$. (Technically, a family of subsets of $V$ forms a "partition," whenever $\cup_{k = 1, ..., c} C_k = V$ and $C_k \cap C_l = \varnothing$, for any $k, l$ in $\{1, 2,..., c\}$.) Then the partition $\mathbf{C}$ is called (non-overlapping) *community structure* of the graph $G$ and each $C_k$ is called a *community,* if a certain benefit function is maximized over this partition $\mathbf{C}$. The most commonly

---

[3] Needless to say that such affiliational networks with attitudinal actors may be formally represented in other ways too: for instance, as restricted tripartite graphs (networks) in the construction of Thomas Fararo and Patrick Doreian (1984).

employed benefit function in community partitioning is the following function $Q$, called *modularity,* which is defined (Newman & Girvan, 2004) as follows:

$Q$ = (fraction of links within communities) - (expected fraction of such links).

In the so-called *null model,* the above expected fraction is calculated on the basis of the hypothetical existence of a random graph, which would preserve the same degree distribution with the examined graph $G$. Hence, the exact expression of modularity becomes (Fortunato, 2010):

$$Q = \sum_{k=1,...,c} [l_k/m - d_k^2/(4m^2)],$$

where $l_k$ is the total number of edges inside community $C_k$ and $d_k$ is the sum of the degrees of all vertices in $C_k$ (in both cases, counting multiplicity of edges). Thus, to obtain a community partitioning one has just to search for that partition, which maximizes the above modularity function $Q$. Note that this optimization problem has been proven to be NP-complete (Brandes *et al.,* 2008) and, therefore, only approximate optimization techniques, such as greedy algorithms, simulated annealing, extremal optimization, expectation maximization, spectral methods etc. can be practically useful.

**Communities formed by the data of the IPPS Survey**

On February 15, 2003, mass protests against the imminent (at that period) war in Iraq took place throughout the world, in which more than seven million people in more than 300 cities all over the globe had participated. This globally sustained mobilization was one of the largest peace protests since the Vietnam War on one single day.[4] On this occasion, an international team of social movement scholars set up a project, called *International Peace Protest Survey* or *IPPS* in abbreviation (2003-4), which was coordinated by Stefaan Walgrave of the University of Antwerp in Belgium (Walgrave & Verhulst, 2009, Diani, 2009).[5] The aim of this project was to study the demonstrations of the global protest event of February 15, 2003, and compare the emergent social movement dynamics in 8 countries: UK, Italy, the Netherlands, Switzerland, USA, Spain, Germany and Belgium. For this purpose, over 10,000 questionnaires were totally distributed and about 6,000 completed questionnaires were sent back, which has made the successful response rate quite satisfactory (well above 50%).

---

[4] For more descriptive information see the Wikipedia article http://en.wikipedia.org/wiki/ February_15,_2003_anti-war_protest.

[5] The survey data bases are accessible at http://webh01.ua.ac.be/m2p/index.php?page= projects&page2=pproject&id=ll.

In particular, in the IPPS survey, participating activists (with varying numbers in each country) were asked to declare their affiliation or involvement with the following 16 (types of) organizations ("affiliational" attributes in the previous terminology): (1) Church, (2) Anti-Racist Organization, (3) Student Organization, (4) Labor Union - Professional Organization, (5) Political Party, (6) Women Organization, (7) Sport - Recreational Organization, (8) Environmental Organization, (9) Art - Music - Educational Org, (10) Neighborhood Organization, (11) Charitable Organization, (12) Anti-Globalist Organization, (13) Third World Organization, (14) Human Rights Organization, (15) Peace Organization, (16) Other Organization.

Furthermore, respondents were asked to answer a number of questions, in which they would express their own opinions and feelings about the meaning, the significance, the reasons and the implications etc. of the war. In particular, activists participating in this event were asked to disclose their positions on the following 10 war-related attitudes (attitudinal attributes in the former terminology): (1) USA Crusade against Islam, (2) Anti-Dictatorial Regime War, (3) UN Security Council Authorized War, (4) War for Oil, (5) Racist War, (6) Iraqi Threat to World Peace, (7) Always Wrong War, (8) War to Overthrow the Iraqi Regime, (9) Feelings against Neoliberal Globalization, (10) Governmental Dissatisfaction.

On the above data set, we have conducted a community analysis resulting the community membership of both "affiliational" and attitudinal attributes which is indicated in the following Table (note that we have disregarded the community membership of respondents, who after all were anonymized in the survey). Each column corresponds to one of the 8 countries where the IPPS survey took place and the numbers in the entries indicate the indices of the community to which the corresponding attribute belongs.

| "Affiliational" Attributes | Italy | Belgium | Switzerland | Germany | Spain | Netherlands | UK | USA |
|---|---|---|---|---|---|---|---|---|
| Church | 1 | 4 | 3 | 1 | 2 | 1 | 1 | 3 |
| Anti-Racist Org | 1 | 1 | 1 | 5 | 2 | 1 | 4 | 2 |
| Student Org | 1 | 6 | 2 | 1 | 2 | 3 | 1 | 5 |
| Labour union - Prof. Org | 2 | 1 | 1 | 2 | 2 | 1 | 2 | 1 |
| Political Party | 1 | 1 | 1 | 2 | 2 | 1 | 4 | 4 |
| Women Org | 1 | 1 | 1 | 5 | 3 | 1 | 4 | 2 |
| Sport - Recr. Org | 1 | 6 | 2 | 1 | 2 | 3 | 1 | 3 |
| Environmental Org | 1 | 1 | 1 | 5 | 2 | 1 | 4 | 2 |
| Art - Music - Educ. Org | 1 | 6 | 4 | 5 | 2 | 1 | 1 | 3 |
| Neighbourhood Org | 1 | 1 | 1 | 5 | 2 | 1 | 4 | 2 |
| Charitable Org | 1 | 1 | 1 | 5 | 2 | 1 | 4 | 2 |
| Anti-Globalist Org | 1 | 1 | 1 | 5 | 2 | 1 | 4 | 2 |
| Third World Org | 1 | 1 | 1 | 5 | 2 | 1 | 4 | 2 |
| Human Rights Org | 1 | 1 | 1 | 5 | 2 | 1 | 4 | 2 |
| Peace Org | 1 | 1 | 1 | 5 | 2 | 1 | 4 | 2 |
| Other Org | 1 | 6 | 3 | 5 | 2 | 1 | 4 | 6 |
| Attitudinal Attributes | | | | | | | | |
| USA Crusade against Islam | 3 | 5 | 2 | 3 | 3 | 2 | 3 | 5 |
| Anti-Dictatorial Regime War | 4 | 2 | 2 | 6 | 1 | 4 | 2 | 6 |
| UN Security Council Authorized War | 4 | 2 | 2 | 6 | 1 | 4 | 2 | 3 |
| War for Oil | 3 | 2 | 2 | 3 | 3 | 2 | 3 | 5 |
| Racist War | 3 | 5 | 2 | 3 | 3 | 2 | 3 | 5 |
| Iraqi Threath to World Peace | 3 | 3 | 2 | 6 | 1 | 2 | 2 | 6 |
| Always Wrong War | 3 | 2 | 2 | 3 | 3 | 2 | 3 | 5 |
| War to Overthrow the Iraqi Regime | 3 | 3 | 2 | 6 | 3 | 2 | 2 | 6 |
| Feelings against Neoliberal Globalization | 3 | 5 | 2 | 3 | 3 | 2 | 3 | 5 |
| Governmental Dissatisfaction | 1 | 2 | 2 | 4 | 3 | 4 | 2 | 6 |
| Computed Modularity | 0.1125 | 0.1003 | 0.0859 | 0.1005 | 0.1284 | 0.0866 | 0.1082 | 0.0836 |

**Table 1**. The communities of affiliations and attributes emerging in the 8 countries of the IPPS survey.

Remarkably, the "mixing" of "affiliational" attributes (organizations) and (actors') attitudes in the community partitioning of the 8 IPPS countries appears to be very low. In 3 countries (Belgium, Germany and the Netherlands), there are no communities, which include both organizations and actors' (respondents') attitudes. This means that, in these three countries, none of the organizations happens to be tight-knit with none of the attitudes. In other 4 countries (Italy, Switzerland, Spain and UK), there is a single community that happens to include both organizations and attitudes. Only in the USA, the mixing is relatively high, as there are three communities with both organizational and attitudinal membership. Thus, given that segregation is the opposite of mixing, what the 8 national IPPS community analyses show is that the macro-link of attributional diversity with community segregation is not locally uniform. Of course, incorporating other (cultural and political) variables into the data set that we have analyzed here could possibly result different patterns of segregation-mixing across national "localities." However, in this way, although the considered constituencies in organizations and attitudes might have changed, the same mechanism-based analytic methodology that we have developed here would still apply. Methodological mechanism-based universality is one of the virtues of analytical sociology when compared to the parochialism of many approaches of current sociology.